\documentclass[conference]{IEEEtran}
\IEEEoverridecommandlockouts
\usepackage{amsmath}
\usepackage{amsfonts}
\usepackage{amssymb}
\usepackage[ruled,vlined]{algorithm2e}
\usepackage{caption}
\usepackage{mathtools}  
\usepackage{amsthm}
\usepackage[]{algorithm2e}
\usepackage{booktabs}
\usepackage{subfigure}
\usepackage{tabulary}
\usepackage{footnote}
\usepackage[export]{adjustbox}
\usepackage{booktabs}
\usepackage[justification=centering]{caption}
\usepackage{comment}

\usepackage{cases}
\usepackage{graphicx}
\usepackage{cite}
\usepackage{multicol}
\usepackage{xcolor}
\usepackage{graphicx}
\graphicspath{{./}{figures/}}
\newcommand\numeq[1]%
  {\stackrel{\scriptscriptstyle(\mkern-1.5mu#1\mkern-1.5mu)}{=}}
\usepackage{stfloats}% <-- added
\usepackage{cuted}
\usepackage{lipsum}  
\usepackage{dsfont}

\usepackage{multicol}  
\usepackage{multirow}  
 
\usepackage{subfigure}
 \usepackage{bm}

\title{Performance Analysis of Semi-Persistent Scheduling Throughput in 5G NR-V2X: A MAC Perspective}

\author{\IEEEauthorblockN{Ran Wei}
\IEEEauthorblockA{{Department of Electrical and Computer Engineering, University of Washington, Seattle, USA}\\
Email: rawe1722@uw.edu} 
}
\begin{document}
% Extremely high throughput (EHT) in IEEE 802.11be for future upgrade of IEEE 802.11 standard focuses on indoor and outdoor wireless local area network (WLAN) throughput maximization.  

\maketitle
\thispagestyle{empty}
\begin{abstract}
The packet throughput in 5th Generation (5G) New Radio (NR) Vehicle-to-Everything (V2X) is highly dependent on the Medium Access Control (MAC) based scheduling algorithm with no base station participation. In particular, the Semi-Persistent Scheduling (SPS) algorithm has been standardized by the 3rd Generation Partnership Project (3GPP) for V2X resource scheduling in the out-of-coverage scenario. This paper analyzes the NR-V2X SPS throughput from the MAC perspective, where the packet reception ratio (PRR) and half-duplex (HD) effect dominate. We first investigate the average  throughput in the fully connected vehicular network, in which all the vehicles share the same throughput. Subsequently, the average throughput as a function of distance in the partially connected vehicular network is analyzed. The Monte Carlo simulation results show that increasing the resource keeping probability can improve the average throughput. Meanwhile, in the partially connected network, the lower resource keeping probability is prone to obtaining the higher throughput gain by increasing the number of subchannels.
\end{abstract}
\section{Introduction} 
\label{introduction}

The rapid development of Vehicle-to-Everything (V2X) communications paves the way for the advanced V2X services such as autonomous driving and remote driving  \cite{sehla2022resource}. In particular, 5G New Radio (NR) V2X, standardized by 3GPP Release 16 \cite{3gpp.22.886}, is gradually becoming the state-of-the-art V2X technology, satisfying different stringent V2X service requirements. One of the key metrics to characterize the quality of V2X service is the average throughout, especially in the out-of-coverage scenario which is quite general in the real world. In such a scenario, each vehicle can only rely on its prior sensing results to schedule resources under a specific scheduling algorithm \cite{cao2020performance}. As a result, the throughput may be severely impacted due to the shortcomings of the sensing mechanism in the decentralized vehicular network. Therefore, it is necessary to investigate the throughput of vehicular networks without the assistance of base stations.

NR-V2X plays as an application of sidelink (SL), which was first introduced in 3GPP Release 12 \cite{3gpp.36.843}, and aimed to enable the device-to-device communications (D2D). Subsequently, Long Term Evolution V2X (LTE-V2X) Mode 3 and Mode 4, as two SL modes, were introduced in Release 14 \cite{release14}. Meanwhile, the LTE-V2X standards also introduced a Semi-Persistent Scheduling (SPS) algorithm which enabled the vehicle to monitor the channel and determine which resources were in use and avoid transmitting on those resources. Recently, it has been evolving as NR-V2X for 5G NR PHY to support advanced services while the SPS algorithm is still adopted in the NR-V2X standard for scheduling resources. The NR-V2X defines two SL modes through the NR-PC5 interface. The Mode 1 signifies the mechanisms that allow direct vehicular communications within gNodeB coverage. The gNodeB in this mode schedules the SL resources for all vehicles. Mode 2 supports direct vehicular communications in the out-of-coverage scenario with a developed SPS mechanism \cite{3gpp.38.885}.

% Meanwhile, some literature also proposed novel scheduling algorithms to provide more flexibility .

 Prior work have investigated the NR-V2X communications from different perspectives \cite{cao2021resource,yi2020enhanced,cao2022optimize,yin2021scheduling,cao2021blockchain,yin2022multihop}. In \cite{cecchini2017ltev2vsim,chen2021performance}, the authors developed a new simulator which focuses on the resource allocation but only depends on limited environmental settings. The authors in \cite{dayal2021adaptive} introduced a new adaptive SPS protocol for decentralized V2X network. The deep reinforcement learning were applied to improve the resource allocation in \cite{cao2021resource,yin2020ns3,zhang2022multiaccess}. In similar literature \cite{ali2021cognitive}, the authors gave a resolution to solve the collision problem under SPS. Both  \cite{8543802} and \cite{8108463} indicated that the parameters in the PHY layer or MAC layer affect the Packet Reception Ratio in C-V2X. Nevertheless, their conclusions were only proved by the simulator without any mathematical model. In addition, a few works focus on relations between the average throughput and the SPS algorithm that become significant if the quality of V2X service is prioritized. Hence, this paper studies the performance analysis of the SPS throughput in the context of NR-V2X Mode 2. The major contributions of this paper are summarized as follows:
\begin{itemize}
\item \textbf{We investigate the average throughput in a quite general V2X scenario, which is out-of-coverage based on the simplified NR-V2X Mode 2}. The average throughput in such a scenario is highly sensitive to the sensing based SPS algorithm due to the shortcomings of the sensing mechanism in the decentralized network.
\item \textbf{We study how the SPS algorithm impacts the average throughput based on our developed analytical model}. The results can provide insights on tuning the average throughput in terms of the SPS parameters to satisfy different advanced V2X service requirements.
\end{itemize}

The rest of this paper is organized as follows. Section \ref{sec:thruput} recaps the channel structure for resource scheduling in NR-V2X and provides the general formula for the average throughput. The analytical models of throughput in both fully connected vehicular networks and partially connected vehicular networks are presented in Section \ref{sec:mac}. In Section \ref{sec:sim}, we conduct Monte Carlo simulations to verify the obtained analytical models. Finally, section \ref{sec:con} draws the conclusions for this paper.

\section{System Architecture}
\label{sec:thruput}

\begin{figure}[t]
    \centering
    \includegraphics[width=.48\textwidth]{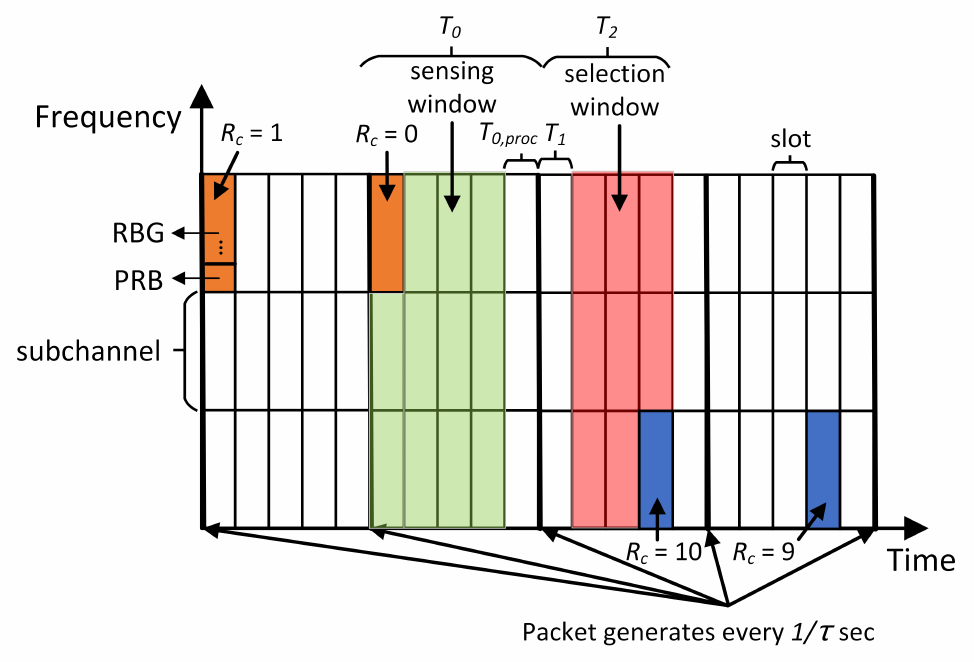}
    \caption{Channel structure for the SL transmission scheduled by the SPS algorithm.}
    \label{fig:SPS}
\end{figure}

% \begin{figure}[t]
%     \centering
%     \includegraphics[width=.45\textwidth]{SPS.png}
%     \caption{The SPS procedure.}
%     \label{fig:SPS}
% \end{figure}
As is shown in Fig. \ref{fig:SPS}, the channel structure consists of physical resource blocks (PRBs) as frequency resources and slots as time resources for NR-V2X packet transmission. A subchannel represents the smallest frequency resource unit for packet transmission/reception, consisting of a group of consecutive PRBs in a slot that form a resource block group (RBG). Each RBG can support the transmission of one packet. Meanwhile, a slot is the smallest unit for scheduling sidelink Basic Safety Message (BSM) package transmissions. Consider that the channel is divided into a (pre-)configured number $n_s$ of contiguous subchannels in the frequency domain, i.e., $n_s$ RBGs in each slot. Each RBG is (pre-)configured to contain the same number of PRBs for packet transmission. Each packet transmission is then scheduled by the sensing-based Semi-Persistent Scheduling (SPS) algorithm\cite{3gpp.36.213,3gpp.36.321}, which is an effective scheduling algorithm to cope with the collision avoidance in V2V communication, especially in the out-of-coverage scenario. 

In the SPS algorithm, as Fig. \ref{fig:SPS} shows, each vehicle randomly selects an available RBG within the selection window according to the results from the sensing window. It then reserves that RBG for a particular duration for consecutive packet transmissions. The number of consecutive transmissions is determined by the re-selection counter (RC). When RC is decremented to 0, it enters its sensing window. The idle RBGs in the sensing window will become available RBGs in the selection window. Then access to selection window, and re-select one of those available RBGs with probability $1 - p_{k}$ as the reserved RBG for the next round consecutive transmissions, where $p_k$ is the resource keeping probability. Otherwise, it will keep the previous reserved RBG for the next round consecutive transmissions. Notice that the packet generation rate also determines the initial RC value $R_c$ that is randomly set within an interval $[5\alpha, 15\alpha]$, where $\alpha = 100/\text{max}(20, 1000/\tau)$ \cite{9345798}. For simplicity, we set the initial $R_c$ to be the average value of such an interval, which characterizes the relationship between the packet generation rate and its corresponding initial RC interval. Therefore, when a vehicle starts a new round of consecutive transmissions, the initial $R_c = 10\alpha$.

We consider a system in which the periodic packet generation rate of each vehicle is $\tau$ ($packets/sec$). Then the average throughput of a vehicle in the vehicular network can be expressed as    
    \begin{equation}
        \Lambda = \tau PRR(1-P_{HD}),
    \end{equation}
where $PRR$ is packet reception ratio of the receiver (RX) vehicle. As the throughput is analyzed under the MAC based scheduling algorithm, we only consider the packet reception from the MAC perspective, i.e., only MAC collision impacts RX's packet reception. Thus $PRR$ is the inverse of the MAC collision probability. $P_{HD}$ is the error probability due to half-duplex (HD) effect. This error occurs when a packet cannot be decoded by a RX vehicle because the RX vehicle is transmitting its own packet in the same slot as the tagged vehicle. The HD effect does not depend on their distance, the scheduling algorithm or the channel occupancy but depends on the packet generation rate, $\tau$, and the number of subframes within a second. Considering 1 ms subframes, the probability of not decoding a packet due to the HD effect can be approximated by the following equation
    \begin{equation}\label{eq:P_HD}
        P_{HD} = \frac{\tau}{1000}.
    \end{equation}

As is mentioned before, The $PRR$ is impacted by the MAC collisions. Assume any packet sent within a given radius (sensing range) can be heard perfectly if it is not interfered by others. Meanwhile, there are some key pre-configures underlying the SPS algorithm should be declared. The packet generation instant of each vehicle is synchronized in terms of packet generation rate. The sensing window size should be equal to $1/\tau$. That is, as Fig \ref{fig:SPS} shows, $T_0$ is at the previous packet generation instant, i.e., $T_0 = 1/\tau$ while $T_{0,proc}$ is pre-configured to be 0. The pre-configure for the selection window size is also equal to $1/\tau$. That is, as is shown in Fig \ref{fig:SPS}, the processing time $T_1$ is pre-configured to be 0, meanwhile, $T_2$ is pre-configured to be the maximum packet delay budget, i.e., $T_2 = 1/\tau$. Therefore, if we denote the slot duration as $t_s (ms)$, the number of RBGs that each vehicle can select for packet transmission in the selection window is
    \begin{equation}\label{eq:N_r}
       N_r = \frac{1000 n_s}{\tau t_s},
    \end{equation}
where $n_s$ is the number of subchannels in the frequency domain. In addition, the key notations are summarized in Table. \ref{tb:notation}.

\begin{table}[h]
\caption{List of key notations.}
\label{tb:notation}
\centering
\fontsize{8}{8}\selectfont{
\begin{tabular}{|c|c|}
\hline
\textbf{Notation}         & \textbf{Definition}  \\ \hline
$p_k$ &  Resource keeping probability \\ \hline
$n_s$ & Number of subchannels \\ \hline
$N_r$ & Number of RBGs in the selection window \\ \hline
$10\alpha$ & Initial RC value \\ \hline
$t_s$ & Slot duration \\ \hline
 $\tau$& Packet generation rate (pkts/sec)\\ \hline
$P_{rs}$ &  Collision probability due to reselection\\ \hline
$P_{kp}$ &  Collision probability due to no reselection\\ \hline
$N_{sen}$ &  Number of vehicles in the sensing range\\ \hline
$N_{a}$ &  Number of available RBGs in the selection window\\ \hline
\end{tabular}
}
\end{table}

\section{Throughput Analysis}
\label{sec:mac}
In this Section, we investigate the average throughput in two types of vehicular networks.
\subsection{Fully connected vehicular network}
\begin{figure}[t]
    \centering
    \includegraphics[width=.3\textwidth]{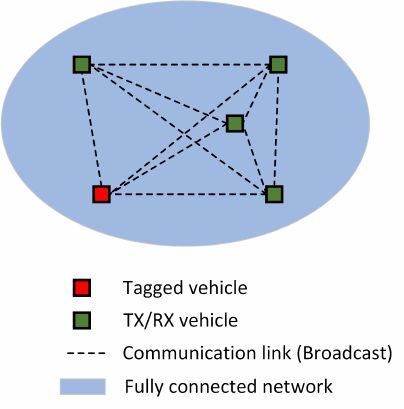}
    \caption{Topology of fully connected vehicular network.}
    \label{fig:fcn}
\end{figure}

We first analyze the throughput in a fully connected vehicular network. As Fig. \ref{fig:fcn} shows, all vehicles can sense each other, i.e., each can transmit to or receive from all the vehicles. The $PRR$ in the fully connected network is impacted by the \textbf{resource selection collisions}. The resource selection collisions occur when at least one vehicle within the sensing range of the tagged vehicle select the same available RBG as the tagged vehicle. If the tagged vehicle senses $N_{sen}$ vehicles, there are $N_{sen}+1$ vehicles in the fully connected network. 
% Fig. \ref{fig:fcn} shows the Markov Chain of RC value in the SPS scheme when $\tau = 10$ (Consider $RRI = 100 ms$). Denote $\pi_i$ is the probability that $RC = i$ where $0\leq i \leq 14$ at the end of the $n^{th}$ RRI (also indicating the beginning of the $(n+1)^{th}$ RRI). In particular, $\pi_0$ is the probability that a vehicle enters the selection window (i.e., $RC = 0$). As a result, $\pi_0$ satisfies 

When a vehicle enters the selection window, it will randomly select an available RBG with probability $1-p_{k}$. Thus the probability that a vehicle selects an available RBG within the same selection window as the tagged vehicle is $(1-p_k)/10\alpha$. For $N_{sen}$ vehicles which are sensed by the tagged vehcile, the probability that $n$ out of $N_{sen}$ vehicles reselect available RBGs within the same selection window as the tagged vehicle is
\begin{equation}\begin{aligned}\label{eq:P_r}
    &Pr\{n \; \text{vehicles reselect}\} \\&= \begin{pmatrix} N_{sen}\\ n \end{pmatrix} \left(\frac{1-p_k}{10\alpha}\right)^n\left(1-\frac{1-p_k}{10\alpha}\right)^{N_{sen}-n}.\end{aligned}
\end{equation}

The expected number of available RBGs depends on $N_r$, $N_{sen}$ and the $PRR$. Assuming each collision happens between two vehicles, the expected number of available RBGs for selection is approximated by
\begin{equation}\label{eq:N_a}
    N_a = N_r - \frac{(1+PRR_{fcn})N_{sen}}{2}.
\end{equation}

If one vehicle needs to select one of available RBG in the same selection window as the tagged vehicle, the probability that both vehicle reselect the same available RBG is $1/N_a$. Hence, given that $n$ vehicles need to reselect available RBGs within the same selection window as the tagged vehicle, the corresponding collision probability is given by
\begin{equation}\label{eq:P_s}
    Pr\{\text{col}|n\;\text{vehicles reselect}\} = 1- \left(1-\frac{1}{N_a}\right)^{n},
\end{equation}
As a result, the collision probability caused by the reselection is given by
\begin{equation}\label{eq:P_c^sr_1}
    P_{rs} = \sum_{n=1}^{N_{sen}}Pr\{\text{col}|n\;\text{vehicles reselect}\}Pr\{n \; \text{vehicles reselect}\}.
\end{equation}
Substituting Eq. (\ref{eq:P_r}) and (\ref{eq:P_s}) into (\ref{eq:P_c^sr_1}), we then obtain
\begin{equation}\label{eq:P_c^sr}
    P_{rs}  = 1- \left[ 1- \frac{1-p_k}{10\alpha N_a}\right]^{N_{sen}},
\end{equation}
where $N_a$ is the number of available RBGs expressed in Eq. (\ref{eq:N_a}). Notice that $P_{rs}$ is a conditional collision probability. As is indicated by SPS, the probability that the tagged vehicle reselect new resources is $1-p_{k}$. On the other hand, given that the tagged vehicle keeps the previous reserved RBG for a new round of consecutive transmissions, a collision can still happen when another vehicle which has collided with the tagged vehicle in the previous round of consecutive transmissions also keeps the same previous reserved RBG. That is, both vehicles still use the previous reserved RBG for the incoming round of consecutive transmissions. Thus, the collision probability, $P_{kp}$, is also a conditional collision probability, given that the tagged vehicle does not reselect new resources. Then both events determine the $PRR$, which can be expressed as
\begin{equation}\begin{aligned}\label{eq:Pc_fcn}
    PRR_{fcn} &= 1-Pr\{\text{reselection}\}Pr\{\text{col}|\text{reselection}\}\\&-Pr\{\text{no reselection}\}Pr\{\text{col}|\text{no reselection}\}\\&= 1 - (1-p_{k})P_{rs} - p_{k}P_{kp}.\end{aligned}
\end{equation}

In Eq. (\ref{eq:Pc_fcn}), the term $P_{kp}$ is hard to obtain if following the SPS procedure, however, it can be represented in terms of $PRR_{fcn}$. For the tagged vehicle, when a collision happens, the probability of such a collision being caused by no reselection is $p_k$. Hence, given $PRR_{fcn}$, $P_{kp}$ can be expressed as $P_{kp} = (1-PRR_{fcn})p_k$. If we substitute $P_{rs}$ and $P_{kp}$ into Eq. (\ref{eq:Pc_fcn}), we obtain
\begin{equation}\label{eq:Pc_fcn_new}
    PRR_{fcn} = \frac{1}{1+p_{k}}\left[p_k+\left(1-\frac{1-p_{k}}{10\alpha N_a}\right)^{N_{sen}}\right].
\end{equation}
Note that each vehicle share the same $PRR$ because they are able to sense each other. This indicates all vehicles have the same average throughput, which can be written as 
$\Lambda_{fcn} = \tau PRR_{fcn}(1-P_{HD})$.

\begin{figure}[t]
    \centering
    \includegraphics[width=.45\textwidth]{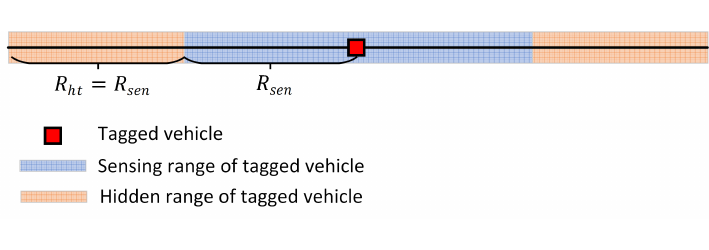}
    \caption{Topology of partially connected vehicular network.}
    \label{fig:pcn}
\end{figure}
\subsection{Partially connected vehicular network}
As Fig.\ref{fig:pcn} shows, different from the fully connected network, the partially connected network considers not only the resource selection collisions but also the \textbf{hidden terminal effects}. The hidden terminal effects occur when at least one vehicle in the hidden terminal region of the tagged vehicle use the same RBG as the tagged vehicle. To model the $PRR$ in the partially connected network, we regard the sensing range of tagged vehicle as the fully connected network when analyzing the resource selection collisions. Such an assumption is reasonable because each vehicle can still sense the same number of vehicles in its sensing range. Notice that in the fully connected network, all vehicles share the same available RBG distribution in the selection window. However, different vehicles in the the partially connected network have different available RBG distributions in the selection window (due to different sensing results), on the other hand, they have the same expected number of available RBGS for selection because of the same traffic load.  
According to Fig.\ref{fig:pcn}, the number of vehicles in the sensing range ($R_{sen}$) of the tagged vehicle should be $N_{sen}=2R_{sen}\rho - 1$,
where $\rho$ is the vehicle density ($vehs/km$).
% Meanwhile, the number of vehicles in the hidden terminal range of the tagged vehicle is $N_{ht} = 2R_{ht}\rho = 2R_{sen}\rho$. 

\begin{figure}[tp]
    \centering
    \includegraphics[width=.35\textwidth]{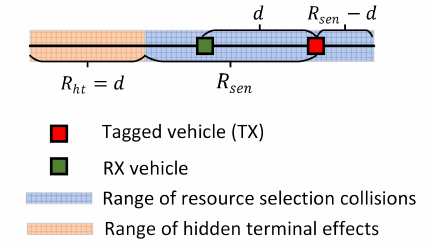}
    \caption{The range of resource selection collisions and hidden terminal effects for a RX vehicle with distance $d \in [0, R_{sen}]$ in the partially connected network.}
    \label{fig:tagged_car_d}
\end{figure}

When the partially connected network is considered, the throughput will be relative to the distance between the tagged TX vehicle and the RX vehicle, as different RX vehicles have different throughput from the tagged vehicle. More precisely, both resource selection collisions and hidden terminal effects depend on the distance. For a specific RX vehicle which is away from the tagged vehicle with distance $d$, the actual $PRR$ of this RX vehicle caused by the resource selection collisions should be higher than the $PRR$ in the case of the fully connected network with the same $N_{sen}$. This is because some vehicles in the sensing range of the tagged vehicle will be out of the sensing range of this RX vehicle. Due to the randomness in RBG selections, the number of packet collisions should be proportional to the distance, as is shown is illustrated in Fig. \ref{fig:tagged_car_d}. Thus, in the partially connected network, the $PRR$ due to the resource selection collisions can be expressed as
\begin{equation}\label{eq:Prr^rsc}
    PRR_{rsc}(d) = 1 - \frac{2R_{sen}-d}{2R_{sen}}\left(1-PRR_{fcn}\right),
\end{equation}
where $PRR_{rsc}(d)$ is a function of the distance $d$, and $PRR_{fcn}$ is expressed in Eq. (\ref{eq:Pc_fcn_new}), where $N_{sen} = 2R_{sen}\rho-1$ now.

We now consider the impact of the hidden terminal effects on the $PRR$. Since the vehicles in the hidden terminal range also uses the same resource selection scheme, they can see the RBGs occupied by vehicles in their own
sensing range. As Fig. \ref{fig:tagged_car_d} shows, for the RX vehicle which is away from the tagged vehicle with distance $d$, the number of hidden vehicles is given by
    \begin{equation}\label{eq:N_ht}
       N_{ht}(d) = R_{ht}\rho = d\rho,
    \end{equation}
where $N_{ht}(d)$ is a function of the distance $d$. Since the hidden vehicles can see at most (on average) $N_{sen}/2$ RBGs used by the vehicles in the sensing range of the tagged vehicle. For simplification, we assume the each hidden vehicle does not use the $N_{sen}/2$ RBGs and randomly select from the remaining $N_r-N_{sen}/2$ RBGs. Therefore, the probability that each hidden vehicle collides with the tagged vehicle is $1/(N_r-N_{sen}/2)$. As there are $N_{ht}(d)$ hidden vehicles, considering both resource selection collisions and the hidden terminal effects, the $PRR$ regarding the distance in the partially connected network is
    \begin{equation}\label{eq:PRR_d}
       PRR_{pcn}(d) = PRR_{rsc}(d)\left(1-\frac{1}{N_r-N_{sen}/2}\right)^{N_{ht}(d)},
    \end{equation}
then the average throughput in the partially connected network is expressed as $\Lambda_{pcn}(d) = \tau PRR_{pcn}(d)(1-P_{HD})$.

\section{Simulation}
\label{sec:sim}
We adapt a Monte Carlo simulation architecture in Matlab of C-V2X networks that were introduced in \cite{cao2021resource} to validate the related analytical models. Table I articulates the main simulation parameters for the fully connected network while the simulation setup in Table II is aimed for the partially connected network. We repeat 40 trials to average the relevant results for each setup.

\begin{figure*}[t]
\begin{minipage}[t]{0.32\linewidth}
\centering
 \subfigure[$\Lambda_{fcn}$ vs $N_{sen}$ ($n_s =5$).]
{\includegraphics[width=2.2in]{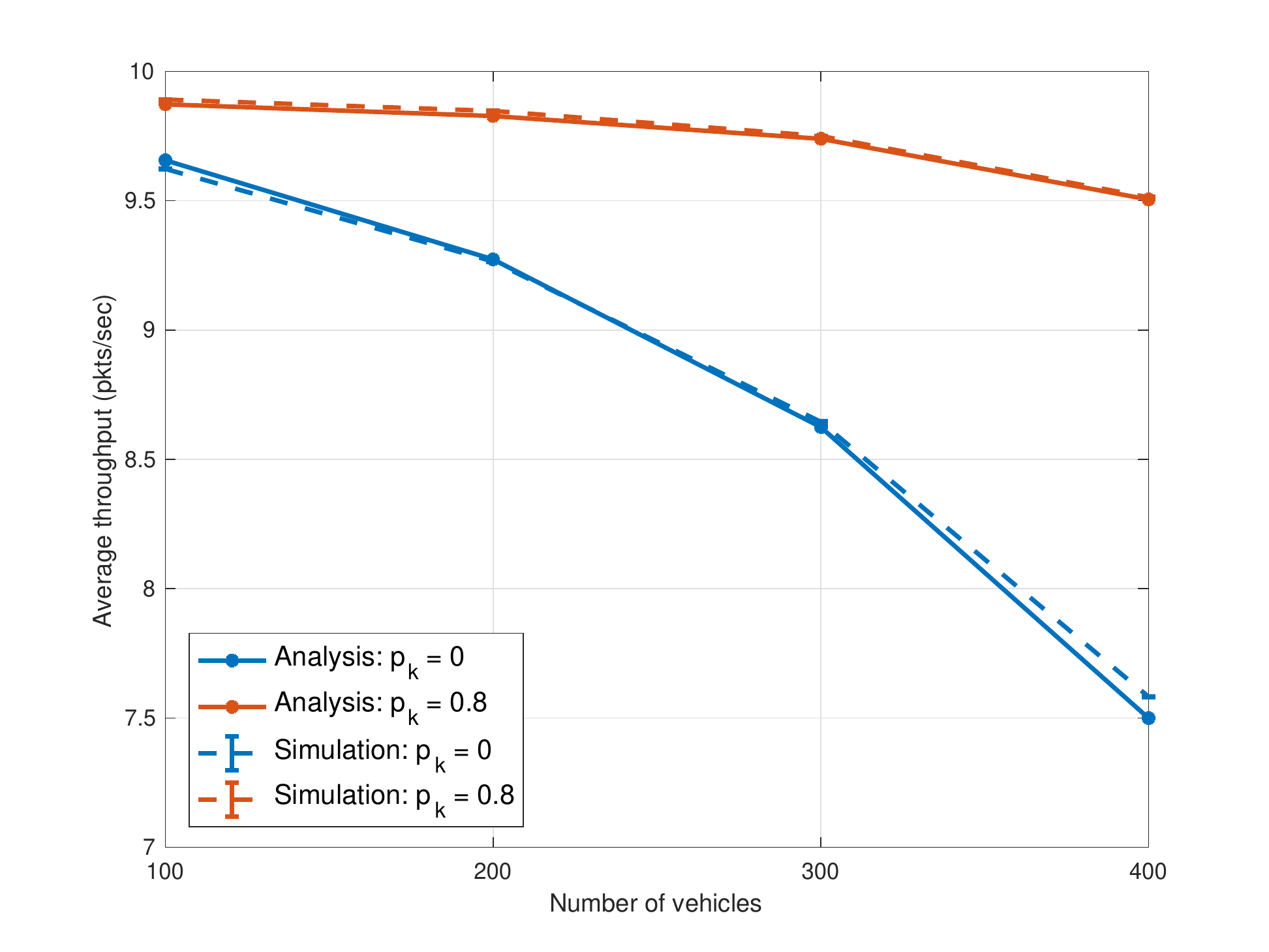}}
\label{fig:6a}
\end{minipage}
\begin{minipage}[t]{0.32\linewidth}
\centering
 \subfigure[$PRR$ vs $p_k$ ($n_s =5$).]
{\includegraphics[width=2.2in]{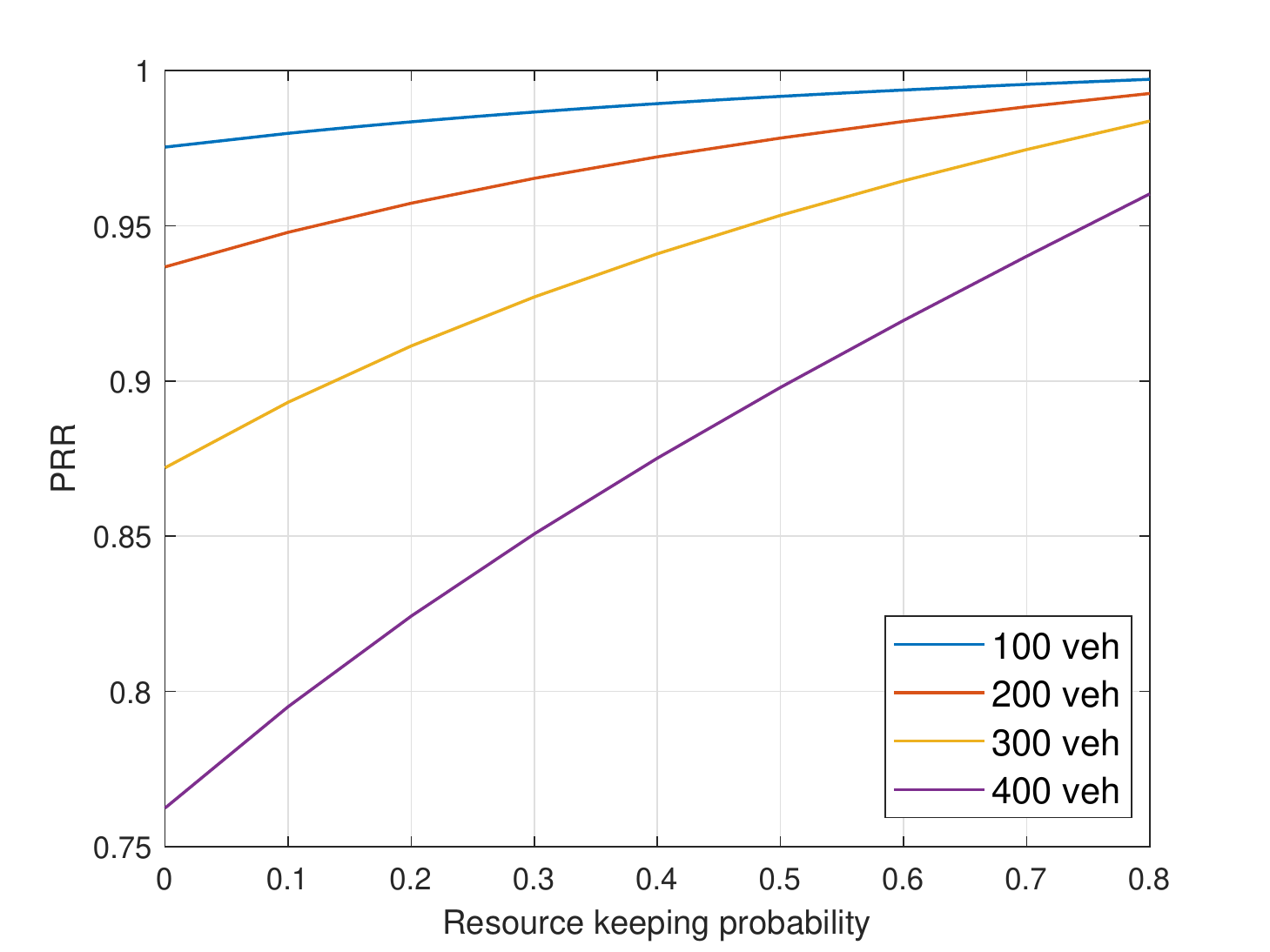}}
\label{fig:6b}
\end{minipage}
\begin{minipage}[t]{0.32\linewidth}
\centering
\subfigure[$\Lambda_{fcn}$ vs $n_s$.]{
\includegraphics[width=2.2in]{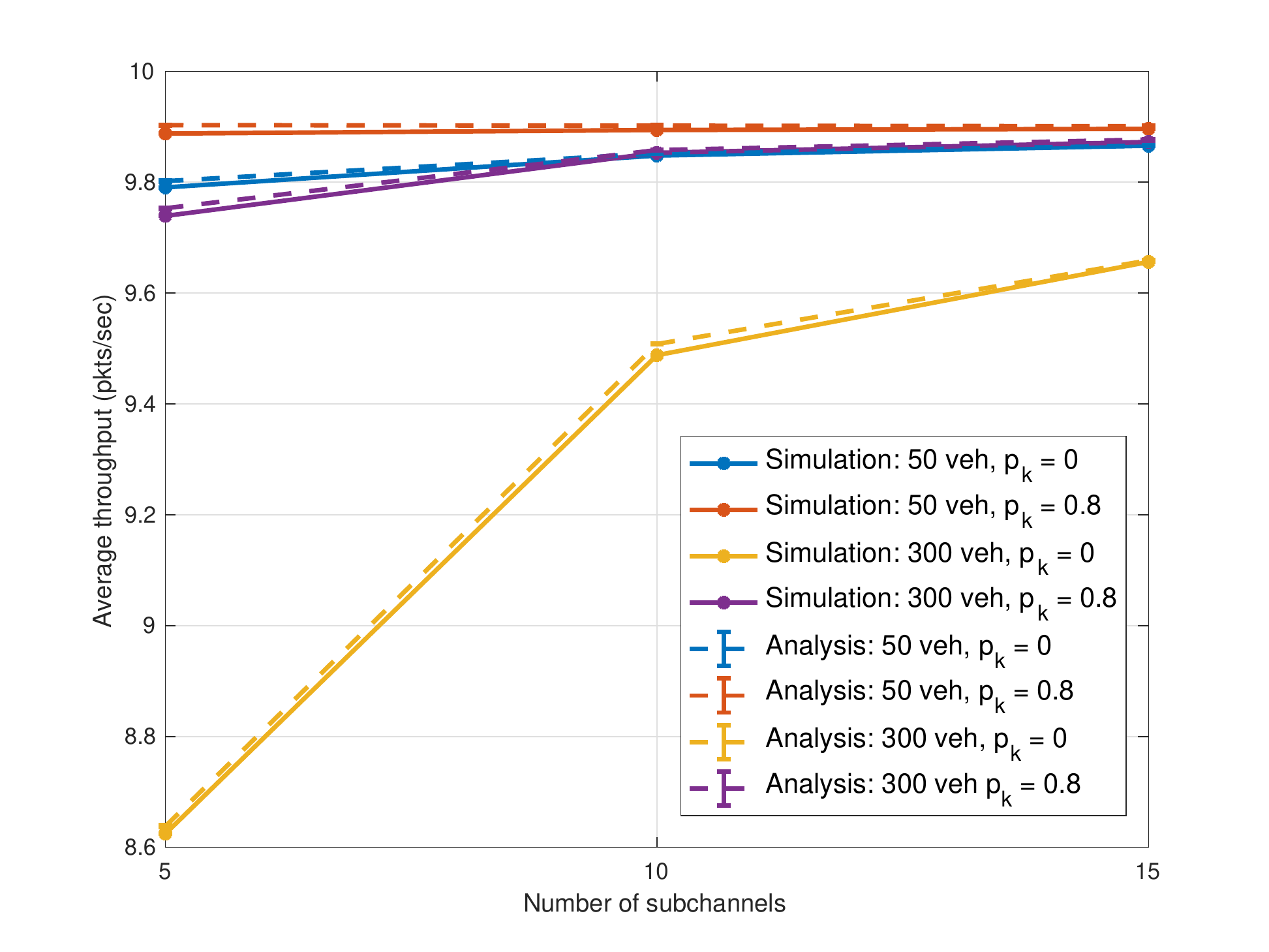}}
\label{fig:6c}
\end{minipage}%
\caption{Average throughput in the fully connected network.}
\label{fig:fully}
\end{figure*}
Fig. \ref{fig:fully} shows the average throughput in the fully connected vehicular network. The analytical model matches quite well with the simulation results. Fig. \ref{fig:fully}(a) shows how the average throughput varies under different total number of vehicles ($N_{sen}$). With a fixed resource keeping probability $p_k$,  when the $N_{sen}$ increases, two vehicles are more likely to select the same RBG for transmission that causes the reception failure, the average throughput thereby decreases. Notice that the average throughput with the resource keeping probability $p_k = 0.8$ is higher than that with $p_k = 0$ whatever the $N_{sen}$ is. Meanwhile, the larger the $N_{sen}$ is, the larger the average throughput difference is. In particular, when $N_{sen} = 400$, the vehicle in the case of $p_k = 0.8$ can receive around two more packets per second than the vehicle in the case of $p_k = 0$. This can be explained by Fig. \ref{fig:fully}(b) that shows how the $PRR$ varies in terms of $p_k$. As we can see, large $N_{sen}$ contributes to larger $PRR$ difference in the range of $p_k$ \footnote{For instance, when $N_{sen} = 400$, the $PRR$ difference between $p_k= 0.8$ and $p_k=0$ is around 0.2, thus the corresponding average throughput difference is around 2 $pkts/sec$, which is consistent with the results in Fig.\ref{fig:fully}(a).}. In addition, while fixing the $N_{sen}$, the $PRR$ increases when $p_k$ increases, leading the average throughput to increase. Therefore, the higher resource keeping probability can guarantee the higher average throughput.

\begin{table}[tp]
\caption{Simulation parameters for the fully connected network.}
\label{tb:simu_main_full}
\centering
\fontsize{8}{8}\selectfont{
\begin{tabular}{|c|c|}
\hline
\textbf{Parameters}         & \textbf{Value}  \\ \hline
Resource keeping probability, $p_k$ & \{0, 0.8\} \\ \hline
Number of vehicles, $N_{sen}$ & \{100, 200, 300, 400\} \\ \hline
Number of sub-channels, $n_s$ &  \{5, 10, 15\} \\ \hline
Slot duration, $t_s$ ($ms$) & 1 \\ \hline
Packet generation rate ($pkts/sec$) & 10 \\ \hline
Simulation time (secs) & 300 \\ \hline
\end{tabular}
}
\end{table}

Fig. \ref{fig:fully}(c) shows the impact of number of subchannels ($n_s$) on the average throughput. When the $N_{sen}$ and $p_k$ are fixed, the average throughput grows with the increasing $n_s$. Since the larger the $n_s$ becomes, the larger the number of RBGs ($N_r$) is, which further increases the expected number of available RBGs ($N_a$) for selection. Hence the corresponding $PRR$ increases, leading the average throughput to increase. It should be noted that the average throughput does not become sensitive to $n_s$ in the case of $N_{sen} = 50$ and $p_k = 0.8$, i.e., the average throughput keeps 9.9 $pkts/sec$ under different $n_s$. In such a case, the HD effect dominates the average throughput, compared with the impact from the MAC collision.

\begin{table}[tp]
\caption{Simulation parameters for the partially connected network.}
\label{tb:simu_main_partial}
\centering
\fontsize{8}{8}\selectfont{
\begin{tabular}{|c|c|}
\hline
\textbf{Parameters}         & \textbf{Value}  \\ \hline
Road length (km) & 5 \\ \hline
Resource keeping probability, $p_k$ &  \{0, 0.8\} \\ \hline
Sensing range (km) & 0.4 \\ \hline
Vehicle density, $\rho$ ($vehs/km$) & \{200\} \\ \hline
Number of sub-channels, $n_s$ & \{5, 10, 15\} \\ \hline
Slot duration, $t_s$ ($ms$) & 1 \\ \hline
Packet generation rate ($pkts/sec$) & \{10\} \\ \hline
Simulation time ($secs$) & 500 \\ \hline
\end{tabular}
}
\end{table}
\begin{figure*}[t]
\begin{minipage}[t]{0.32\linewidth}
\centering
 \subfigure[$\Lambda_{pcn}(d)$ vs $d$ ($n_s=5$).]
{\includegraphics[width=2.2in]{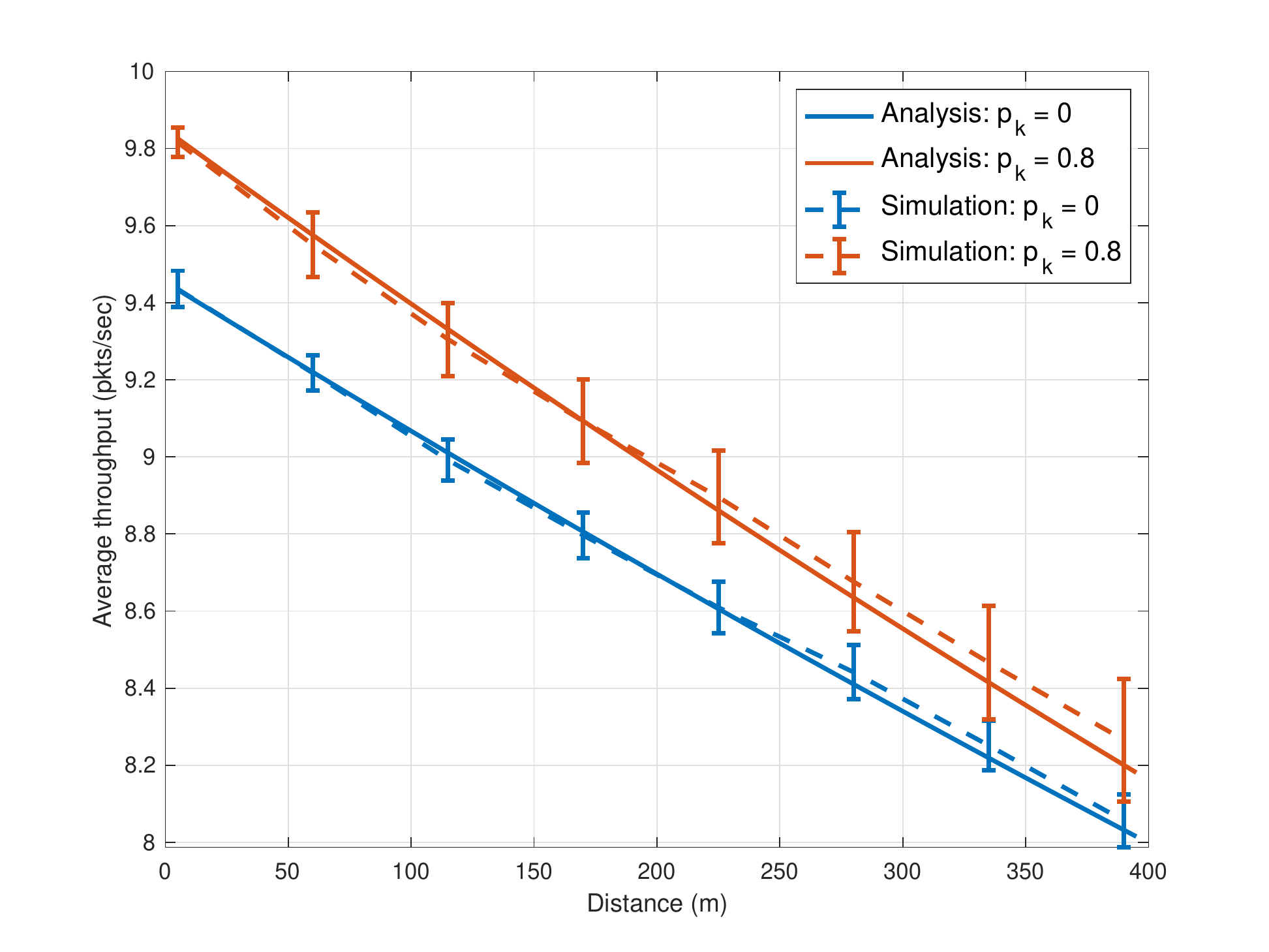}}
\label{fig:7a}
\end{minipage}
\begin{minipage}[t]{0.32\linewidth}
\centering
\subfigure[$\Lambda_{pcn}(d)$ vs $d$ ($p_k = 0$).]{
\includegraphics[width=2.2in]{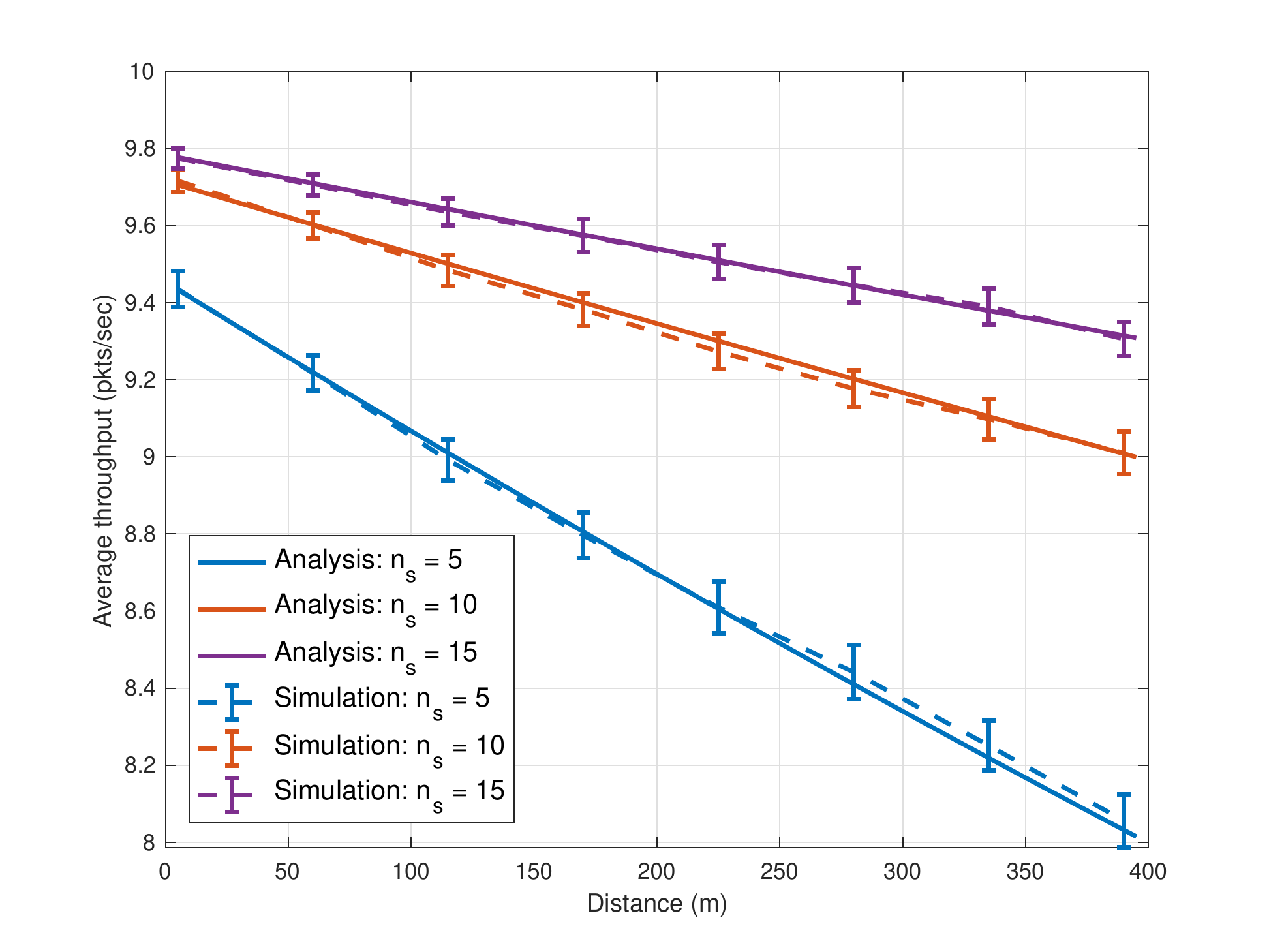}}
\label{fig:7b}
\end{minipage}%
\begin{minipage}[t]{0.32\linewidth}
\centering
\subfigure[Average vehicular network throughput vs $n_s$.]{
\includegraphics[width=2.2in]{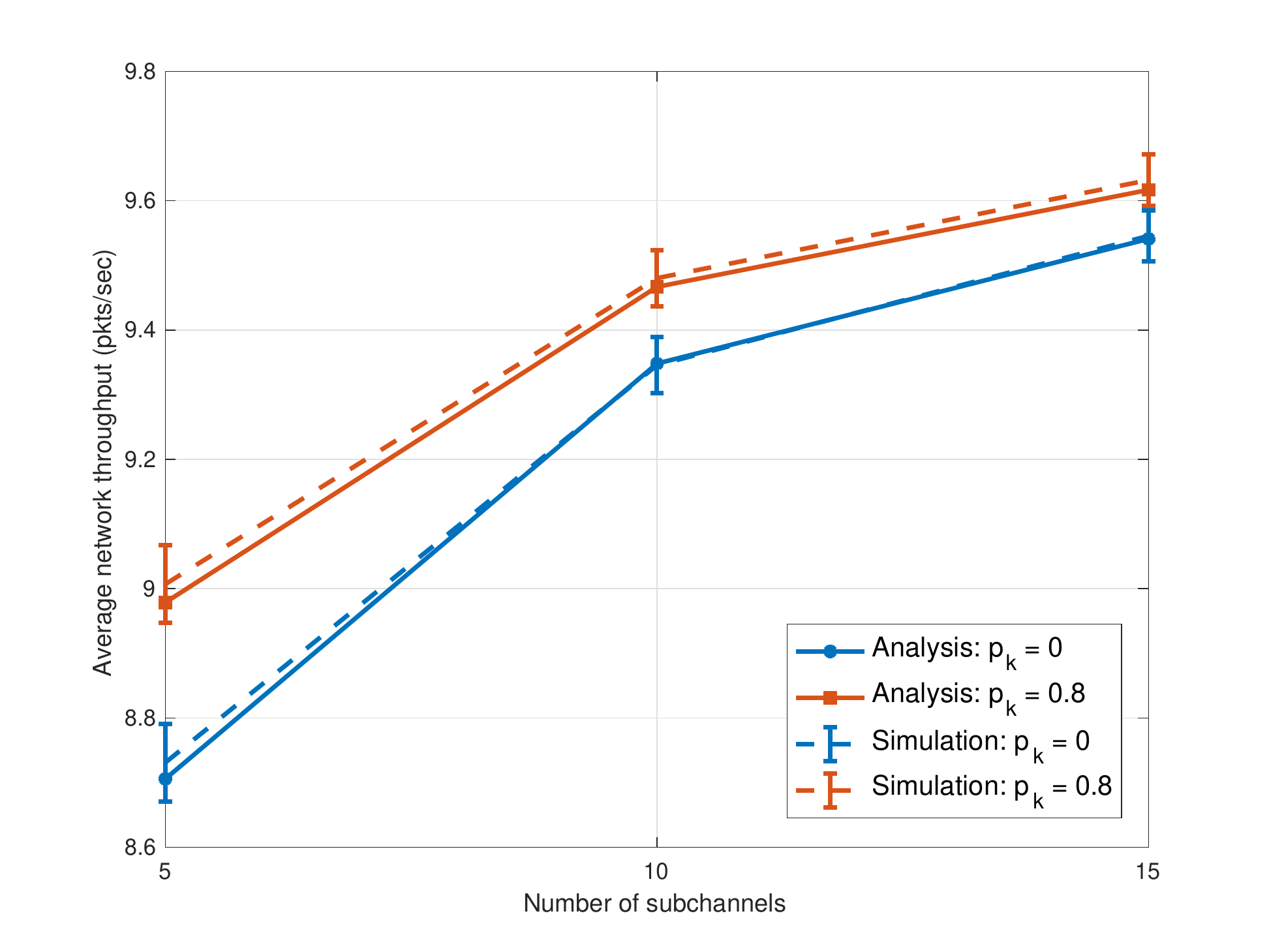}}
\label{fig:7c}
\end{minipage}
\caption{Average throughput in the partially connected network under 200 vehicles/km.}
\label{fig:partially}
\end{figure*}

To investigate the average throughput regarding the distance in the partially connected network, we fixed the vehicle density. Fig. \ref{fig:partially} shows the average throughput regarding the distance under 200 vehicles/km. The analytical model also matches well with the simulation results under all setup. As Fig. \ref{fig:partially}(a) shows, with a fixed number of subchannels ($n_s$), the throughput with $p_k = 0.8$ is always higher than the throughput with $p_k = 0$ whatever the $d$ is. However, the larger the $d$ is, the smaller the throughput difference is. Such throughput difference can be explained by Eq. (\ref{eq:Prr^rsc}), where the larger the $d$ is, the less impact that the resource selection collisions has on the $PRR$ of the RX vehicle. Therefore, under different $p_k$, the average throughput becomes close to each other with the increasing $d$.  Moreover, with a fixed resource keeping probability, as is shown in \ref{fig:partially}(b), the average throughput can be improved by raising $n_s$. Since the hidden terminal effects is relative to $n_s$ which determines $N_r$, as Eq. (\ref{eq:PRR_d}) shows, increasing $n_s$ will mitigate the hidden terminal effects by decreasing the probability that each hidden vehicle collides with the tagged vehicle, the average throughput then increases. Hence, among three cases, the larger the $d$ (i.e., more hidden vehicles are included) is, the larger the average throughput difference is. 

Furthermore, we take the average of the average throughput ($\Lambda_{pcn}(d)$) in terms of all investigated distances to characterize  vehicular network throughput \footnote{The vehicular network throughput is investigated in terms of the vehicles which are not impacted by the edge effects.}. As Fig. \ref{fig:partially}(c) shows, the average network throughput grows with the increasing $n_s$ in both cases. This is due to that increasing $n_s$ mitigates not only the resource selection collisions but also the hidden terminal effects. In addition, the average network throughput with $p_k = 0.8$ is always higher than that with $p_k = 0$ regardless of the number of subchannels. However, such a difference decreases with the increasing $n_s$. The main reason is that the average network throughput with $p_k = 0$ is quite more sensitive to $n_s$ than that with $p_k = 0.8$. Therefore, the lower resource keeping probability can obtain the higher throughput gain by increasing the number of subchannels.
% The decrements of throughput under the low number of subchannels is more significant than that under the large number of subchannels 
% ,i.e., the packet collision with $p_k = 0$ happens far more frequently than that with $p_k = 0.8$ under under the low number of subchannels, while both case 
\section{Conclusion}
\label{sec:con}
In this paper, we focused on the throughput performance of Semi-Persistent Scheduling (SPS) in NR-V2X. From the MAC perspective, the average throughput was mainly impacted by the packet reception ratio ($PRR$) and the half-duplex (HD) effect. We first analyzed the average throughput in the fully connected network, where the $PRR$ was only impacted by resource selection collisions. We then investigated the packet throughput in the partially connected network, where the resource selection collisions and hidden terminal effects which determined the $PRR$ were both relative to the distance between the transmitter and receiver vehicle. Finally, the analytical models were verified through the corresponding simulation results. The numerical results show that increasing the resource keeping probability can improve the average throughput in both investigated networks. In addition, in the partially connected network, the lower resource keeping probability helps obtain the higher throughput gain with by increasing the number of subchannels. 
%\input{conclusion}

% \appendix

% \input{appendix}

\bibliographystyle{IEEEtran}
\bibliography{reference}

% Generated by IEEEtran.bst, version: 1.14 (2015/08/26)
\begin{thebibliography}{10}
\providecommand{\url}[1]{#1}
\csname url@samestyle\endcsname
\providecommand{\newblock}{\relax}
\providecommand{\bibinfo}[2]{#2}
\providecommand{\BIBentrySTDinterwordspacing}{\spaceskip=0pt\relax}
\providecommand{\BIBentryALTinterwordstretchfactor}{4}
\providecommand{\BIBentryALTinterwordspacing}{\spaceskip=\fontdimen2\font plus
\BIBentryALTinterwordstretchfactor\fontdimen3\font minus
  \fontdimen4\font\relax}
\providecommand{\BIBforeignlanguage}[2]{{%
\expandafter\ifx\csname l@#1\endcsname\relax
\typeout{** WARNING: IEEEtran.bst: No hyphenation pattern has been}%
\typeout{** loaded for the language `#1'. Using the pattern for}%
\typeout{** the default language instead.}%
\else
\language=\csname l@#1\endcsname
\fi
#2}}
\providecommand{\BIBdecl}{\relax}
\BIBdecl

\bibitem{sehla2022resource}
K.~Sehla, T.~M.~T. Nguyen, G.~Pujolle, and P.~B. Velloso, ``Resource allocation
  modes in {C-V2X}: From {LTE-V2X} to {5G-V2X},'' \emph{IEEE Internet of Things
  Journal}, 2022.

\bibitem{3gpp.22.886}
3GPP, ``Study on enhancement of {3GPP} support for {5G} {V2X} services,'' The
  3rd Generation Partnership Project {(3GPP)}, Tech. Rep. {TR}22.886, Dec 2018.

\bibitem{cao2020performance}
L.~Cao, H.~Yin, J.~Hu, and L.~Zhang, ``Performance analysis and improvement on
  {DSRC} application for {V2V} communication,'' in \emph{2020 IEEE 92nd
  Vehicular Technology Conference (VTC2020-Fall)}.\hskip 1em plus 0.5em minus
  0.4em\relax IEEE, 2020, pp. 1--6.

\bibitem{3gpp.36.843}
3GPP, ``Study on {LTE} device to device proximity services; radio aspects,''
  The 3rd Generation Partnership Project {(3GPP)}, Tech. Rep. {TR}36.843, March
  2014.

\bibitem{release14}
------, ``Evolved universal terrestrial radio access ({E-UTRA}) and evolved
  universal terrestrial radio access network ({E-UTRAN}),'' The 3rd Generation
  Partnership Project {(3GPP)}, Tech. Rep. {TR}36.300, Jun 2017.

\bibitem{3gpp.38.885}
------, ``{S}tudy on {NR} {V}ehicle-to-{E}verything {(V2X)},'' The 3rd
  Generation Partnership Project {(3GPP)}, Tech. Rep. {TR}38.885, Mar 2019.

\bibitem{cao2021resource}
L.~Cao and H.~Yin, ``Resource allocation for vehicle platooning in {5G NR-V2X}
  via deep reinforcement learning,'' in \emph{2021 IEEE International Black Sea
  Conference on Communications and Networking (BlackSeaCom)}.\hskip 1em plus
  0.5em minus 0.4em\relax IEEE, 2021, pp. 1--7.

\bibitem{yi2020enhanced}
S.~Yi, G.~Sun, and X.~Wang, ``Enhanced resource allocation for {5G V2X} in
  congested smart intersection,'' in \emph{2020 IEEE 92nd Vehicular Technology
  Conference (VTC2020-Fall)}.\hskip 1em plus 0.5em minus 0.4em\relax IEEE,
  2020, pp. 1--5.

\bibitem{cao2022optimize}
L.~Cao, H.~Yin, R.~Wei, and L.~Zhang, ``Optimize semi-persistent scheduling in
  {NR-V2X}: An age-of-information perspective,'' in \emph{2022 IEEE Wireless
  Communications and Networking Conference (WCNC)}.\hskip 1em plus 0.5em minus
  0.4em\relax IEEE, 2022, pp. 2053--2058.

\bibitem{yin2021scheduling}
H.~Yin, L.~Cao, and X.~Deng, ``Scheduling and resource allocation for multi-hop
  {URLLC} network in {5G} sidelink,'' in \emph{2021 IEEE 94th Vehicular
  Technology Conference (VTC2021-Fall)}.\hskip 1em plus 0.5em minus 0.4em\relax
  IEEE, 2021, pp. 1--7.

\bibitem{cao2021blockchain}
L.~Cao and H.~Yin, ``A blockchain-empowered platoon communication scheme for
  vehicular safety applications,'' in \emph{2021 IEEE 94th Vehicular Technology
  Conference (VTC2021-Fall)}.\hskip 1em plus 0.5em minus 0.4em\relax IEEE,
  2021, pp. 1--6.

\bibitem{yin2022multihop}
H.~Yin, S.~Roy, and L.~Cao, ``Routing and resource allocation for {IAB}
  multi-hop network in {5G} advanced,'' \emph{IEEE Transactions on
  Communications}, pp. 1--1, 2022.

\bibitem{cecchini2017ltev2vsim}
G.~Cecchini, A.~Bazzi, B.~M. Masini, and A.~Zanella, ``{LTEV2Vsim}: An
  {LTE-V2V} simulator for the investigation of resource allocation for
  cooperative awareness,'' in \emph{2017 5th IEEE International Conference on
  Models and Technologies for Intelligent Transportation Systems
  (MT-ITS)}.\hskip 1em plus 0.5em minus 0.4em\relax IEEE, 2017, pp. 80--85.

\bibitem{chen2021performance}
M.~Chen, R.~Chai, H.~Hu, W.~Jiang, and L.~He, ``Performance evaluation of
  {C-V2X} mode 4 communications,'' in \emph{2021 IEEE Wireless Communications
  and Networking Conference (WCNC)}.\hskip 1em plus 0.5em minus 0.4em\relax
  IEEE, 2021, pp. 1--6.

\bibitem{dayal2021adaptive}
A.~Dayal, V.~K. Shah, B.~Choudhury, V.~Marojevic, C.~Dietrich, and J.~H. Reed,
  ``Adaptive semi-persistent scheduling for enhanced on-road safety in
  decentralized {V2X} networks,'' in \emph{2021 IFIP Networking Conference
  (IFIP Networking)}.\hskip 1em plus 0.5em minus 0.4em\relax IEEE, 2021, pp.
  1--9.

\bibitem{yin2020ns3}
H.~Yin, P.~Liu, K.~Liu, L.~Cao, L.~Zhang, Y.~Gao, and X.~Hei, ``ns3-ai:
  Fostering artificial intelligence algorithms for networking research,'' in
  \emph{Proceedings of the 2020 Workshop on Ns-3}, 2020, pp. 57--64.

\bibitem{zhang2022multiaccess}
L.~Zhang, H.~Yin, S.~Roy, and L.~Cao, ``Multiaccess point coordination for
  next-gen {Wi-Fi} networks aided by deep reinforcement learning,'' \emph{IEEE
  Systems Journal}, 2022.

\bibitem{ali2021cognitive}
M.~Ali and Y.-T. Kim, ``Cognitive collision resolution for enhanced performance
  in {C-V2X} sidelink mode 4,'' in \emph{2021 22nd Asia-Pacific Network
  Operations and Management Symposium (APNOMS)}.\hskip 1em plus 0.5em minus
  0.4em\relax IEEE, 2021, pp. 102--107.

\bibitem{8543802}
A.~Bazzi, G.~Cecchini, A.~Zanella, and B.~M. Masini, ``Study of the impact of
  {PHY} and {MAC} parameters in 3gpp {C-V2V} mode 4,'' \emph{IEEE Access},
  vol.~6, pp. 71\,685--71\,698, 2018.

\bibitem{8108463}
R.~Molina-Masegosa and J.~Gozalvez, ``System level evaluation of {LTE-V2V} mode
  4 communications and its distributed scheduling,'' in \emph{2017 IEEE 85th
  Vehicular Technology Conference (VTC Spring)}, 2017, pp. 1--5.

\bibitem{3gpp.36.213}
3GPP, ``Evolved universal terrestrial radio access {(E-UTRA)}; physical layer
  procedures,'' The 3rd Generation Partnership Project {(3GPP)}, Tech. Rep.
  {TR}36.213, Dec 2020.

\bibitem{3gpp.36.321}
------, ``Evolved universal terrestrial radio access {(E-UTRA)}; medium access
  control {(MAC)} protocol specification,'' The 3rd Generation Partnership
  Project {(3GPP)}, Tech. Rep. {TR}36.321, Dec 2020.

\bibitem{9345798}
M.~H.~C. Garcia, A.~Molina-Galan, M.~Boban, J.~Gozalvez, B.~Coll-Perales,
  T.~Şahin, and A.~Kousaridas, ``A tutorial on {5G NR V2X} communications,''
  \emph{IEEE Communications Surveys Tutorials}, vol.~23, no.~3, pp. 1972--2026,
  2021.

\end{thebibliography}
\end{document}